\documentclass[11pt]{article}
\usepackage[margin=1in]{geometry}
\usepackage{amsmath,amssymb}
\usepackage{graphicx}
\usepackage{booktabs}
\usepackage{array}
\usepackage{microtype}
\usepackage[hidelinks]{hyperref}
\usepackage{url}
\newcommand{\TtwoStar}{\ensuremath{T_2^{*}}}
\newcommand{\RtwoStar}{\ensuremath{R_2^{*}}}
\newcommand{\Racc}{\ensuremath{R}}
\title{Quantitative \TtwoStar{} Fidelity in Accelerated Radial\\
Ultrashort-Echo-Time MRI of the Lumbar Cartilage Endplate\\[0.6em]
\large A synthetic computational validation study}
\author{Yousef Riyazifar\\
Independent Researcher\\
Karlsruhe, Germany\\
\href{mailto:y.riyazifar@gmail.com}{y.riyazifar@gmail.com} \;|\; \href{https://orcid.org/0009-0007-6548-8800}{ORCID: 0009-0007-6548-8800}}
\date{}
\begin{document}
\maketitle
\begin{center}
\small Manuscript draft for scientific review. All experimental results reported here are synthetic; no patient, volunteer, physical-phantom, or scanner raw data were used in the final validation.
\end{center}
\clearpage

\section*{Simple Summary}
The cartilage endplate is a thin tissue layer between the vertebral body and the intervertebral disc. Its short transverse relaxation makes conventional MRI insensitive to part of its signal, whereas ultrashort-echo-time (UTE) MRI can preserve this rapidly decaying component. Quantitative \TtwoStar{} mapping may provide information about endplate composition, but accelerating a UTE acquisition changes the reconstruction problem and may alter the measured biomarker even when the reconstructed image appears sharp. We created a fully synthetic multi-coil radial UTE framework of the lumbar spine and compared conjugate-gradient SENSE, joint multi-echo FISTA, and direct model-based \TtwoStar{} reconstruction across increasing undersampling. In a final frozen Monte Carlo experiment, the direct model was accurate at low and moderate acceleration, while severe undersampling produced systematic \TtwoStar{} underestimation. Joint-FISTA produced the lowest image error but not the best quantitative \TtwoStar{} fidelity. These findings show why accelerated quantitative MRI should be evaluated using the final biomarker, not image appearance alone, and define a testable framework for subsequent validation with physical phantoms and Siemens 3-T raw data.

\begin{abstract}
\textbf{Background/Objectives:} The lumbar cartilage endplate (CEP) is a short-\TtwoStar{} structure whose composition has been linked to disc health. UTE MRI enables direct CEP visualization, but acceleration and reconstruction may alter quantitative \TtwoStar{} even when image quality improves. This study evaluated how radial undersampling and reconstruction strategy affect CEP \TtwoStar{} fidelity in a controlled synthetic setting.

\textbf{Methods:} A $96 \times 96$ sagittal lumbar phantom with five CEP levels (ground-truth \TtwoStar{}: 24, 22, 20, 18, and 16 ms) was simulated with six UTE echoes (TE, 0.244--25.2 ms), eight receive coils, and radial sampling. Fully sampled reference data used 302 spokes; retrospective accelerations $\Racc=1,2,4,6,$ and 8 used 302, 151, 76, 50, and 38 spokes, respectively. Reconstructions were CG-SENSE, frozen joint multi-echo FISTA, and a direct nonlinear model estimating $\rho$ and \RtwoStar{} from k-space. Reconstruction settings were frozen before final validation. Twenty independent complex-noise realizations were evaluated per acceleration. Primary endpoints were CEP \TtwoStar{} RMSE, bias, within-CEP dispersion, echo-decay fidelity, and 95\% bootstrap confidence intervals.

\textbf{Results:} Direct model-map RMSE increased from 0.172 ms (95\% CI, 0.153--0.191) at $\Racc=1$ to 0.800 ms (0.766--0.836) at $\Racc=4$, 1.472 ms (1.386--1.557) at $\Racc=6$, and 2.256 ms (2.182--2.331) at $\Racc=8$. At $\Racc=4$, joint-FISTA had lower image NRMSE than the direct model (0.132 vs. 0.203), whereas the direct map had lower level-wise \TtwoStar{} RMSE than CG-SENSE and joint-FISTA ROI fits (0.800 vs. 1.448 and 1.288 ms; estimators differed). At $\Racc=8$, CG-SENSE ROI fitting had the lowest \TtwoStar{} RMSE (1.501 ms), whereas the direct model showed systematic underestimation (bias, $-2.136$ ms; concordance correlation coefficient, 0.723; Bland--Altman limits, $-3.605$ to $-0.668$ ms).

\textbf{Conclusions:} Image fidelity and quantitative biomarker fidelity diverged under radial acceleration. In this synthetic framework, $\Racc=4$ represented a promising operating point for direct CEP \TtwoStar{} mapping, whereas $\Racc=8$ produced substantial bias and spatial instability. Scanner, phantom, and in vivo validation are required before clinical translation.
\end{abstract}

\noindent\textbf{Keywords:} ultrashort echo time; UTE; cartilage endplate; lumbar spine; \TtwoStar{} mapping; radial MRI; quantitative MRI; model-based reconstruction; compressed sensing; Monte Carlo validation

\section{Introduction}
The cartilaginous endplate is a thin interface between vertebral bone and the intervertebral disc and contributes to solute transport between the vertebral marrow and disc. Conventional spin-echo MRI provides limited direct signal from short-$T_2$ components of the endplate, while ultrashort-echo-time (UTE) methods can visualize structures that decay before conventional echo times \cite{bae2013}. Subsequent work established that CEP \TtwoStar{} measured with very short echo times is sensitive to biochemical composition, including glycosaminoglycan and water content, although the measurement is strongly orientation dependent \cite{fields2015}.

In vivo quantitative studies have further supported the biological relevance of CEP \TtwoStar{}. Wang et al. reported spatial variation and repeatability of lumbar CEP \TtwoStar{} and observed associations between lower CEP \TtwoStar{} and more advanced adjacent disc degeneration in younger participants \cite{wang2021}. In patients with chronic low back pain, CEP composition remained associated with adjacent nucleus-pulposus composition after adjustment for demographic factors \cite{bonnheim2022}. UTE morphology has also been associated with disc degeneration in cadaveric and patient studies \cite{finkenstaedt2023,ji2023}. More recently, dual-inversion-recovery-prepared UTE has enabled high-contrast CEP depiction in the spine at 3 T \cite{athertya2024}.

A major obstacle to quantitative UTE MRI is acquisition time. Radial sampling is attractive for short-TE imaging, but acceleration reduces the number of spokes supporting each echo and increases streaking, noise amplification, and the risk that reconstruction alters the relative signal amplitudes across echo times. This distinction is especially important for relaxometry: a reconstruction can produce a visually sharp image while changing the decay slope that determines \TtwoStar{}. The problem is not unique to UTE. Model-based quantitative MRI methods have shown that parameter maps can be estimated directly from undersampled raw data by incorporating the MR signal model into the inverse problem \cite{block2009,sumpf2011,sumpf2014}.

The present study was designed as a synthetic stress test rather than a clinical validation. We asked whether image-domain and model-based reconstruction strategies preserve a predefined lumbar CEP \TtwoStar{} biomarker as radial undersampling increases. The reconstruction-development phase was completed before a final frozen Monte Carlo validation. Our primary objective was to characterize quantitative \TtwoStar{} fidelity across acceleration factors and to test whether image fidelity and biomarker fidelity provide concordant rankings of reconstruction methods.

\section{Materials and Methods}
\subsection{Study Design and Synthetic Ground Truth}
This was a fully synthetic computational study. No patients, volunteers, animals, physical phantoms, or scanner raw data were used; institutional review board approval and informed consent were therefore not applicable. The final validation was intentionally separated from reconstruction development: all reconstruction and regularization parameters were frozen before the V4.4 Monte Carlo experiment, noise seeds were fixed prospectively, and no tuning was performed on the final validation realizations.

A stylized sagittal lumbar phantom ($96 \times 96$ matrix) represented five motion segments from L1--L2 through L5--S1 and included CEP, nucleus pulposus, annulus fibrosus, vertebral marrow, cortical/subchondral bone, and paraspinal muscle compartments. Five level-specific CEP \TtwoStar{} values (24, 22, 20, 18, and 16 ms from L1--L2 to L5--S1) spanned the literature-anchored in vivo range reported for higher- and lower-\TtwoStar{} CEP composition-like states \cite{wang2021}. These values were not used as diagnostic categories. CEP orientation relative to $B_0$ was held fixed because the magic-angle dependence of CEP \TtwoStar{} can be as large as the biological differences of interest \cite{fields2015}.

\subsection{UTE Signal and Multi-Coil Radial Acquisition Model}
The synthetic UTE timing was based on an in vivo multi-echo UTE-Cones protocol reported for CEP composition imaging: repetition time 32 ms, flip angle $19^{\circ}$, and six echo times of 0.244, 5.2, 10.2, 15.2, 20.2, and 25.2 ms \cite{wang2021}. Tissue-dependent proton-density/$T_1$ amplitude factors were combined with mono-exponential \TtwoStar{} decay. For the direct reconstruction, all echo-independent amplitude terms were absorbed into a complex $\rho$ map, so that the predicted echo signal was modeled as
\begin{equation}
S_e(\mathbf{r}) = \rho(\mathbf{r})\exp\!\left[-\mathrm{TE}_e R_2^{*}(\mathbf{r})\right],
\qquad
T_2^{*}(\mathbf{r}) = \frac{1}{R_2^{*}(\mathbf{r})}.
\end{equation}

Eight smooth synthetic receive-coil sensitivity maps were used. A matched radial forward/adjoint operator generated complex multi-coil k-space. The dense reference used 302 spokes. Acceleration factors $\Racc=1,2,4,6,$ and 8 corresponded to 302, 151, 76, 50, and 38 spokes, respectively. Complex Gaussian k-space noise was added at the fixed final normalized setting ($\sigma \approx 0.004975$ per complex channel). This noise level was selected for the computational experiment and was not calibrated to reproduce clinical repeatability.

\subsection{Reconstruction Methods}
\paragraph{CG-SENSE.} Multi-coil least-squares reconstruction was performed using an unregularized conjugate-gradient sensitivity-encoding formulation. This served as a conventional data-consistent iterative comparator \cite{pruessmann1999}.

\paragraph{Joint multi-echo FISTA.} A data-consistent FISTA reconstruction used a shared Haar-wavelet group-sparsity prior across the echo dimension. The relative threshold ($5 \times 10^{-5}$) had been selected in a separate development/calibration experiment and was frozen before the final Monte Carlo validation. FISTA was included because sparse regularization can substantially improve image fidelity under undersampling while potentially changing relative echo amplitudes \cite{beck2009,lustig2007}.

\paragraph{Direct model-based \TtwoStar{} reconstruction.} A nonlinear inverse formulation jointly estimated the complex $\rho$ map and the real \RtwoStar{} map from all echo k-space data. Analytical gradients propagated data residuals through the exponential signal model. Earlier development experiments tested wavelet regularization of $\rho$ and edge-aware spatial regularization of \RtwoStar{}; both calibration procedures selected zero regularization. Accordingly, the final model used $\rho$ regularization $=0$ and \RtwoStar{} edge-aware $\beta=0$ with 75 nonlinear iterations. This choice avoided introducing a smoother but biased map after calibration showed that increased parameter regularization reduced voxel variance at the cost of greater \TtwoStar{} error. Direct parameter estimation follows the broader model-based quantitative MRI framework established for accelerated relaxometry \cite{block2009,sumpf2011,sumpf2014}.

\subsection{Quantitative Endpoints}
For CG-SENSE and joint-FISTA, signal was averaged within each predefined synthetic CEP level at each echo and fitted to a mono-exponential decay, producing a regional ROI-fit \TtwoStar{}. The direct model additionally produced a \TtwoStar{} parameter map; the mean \TtwoStar{} within each CEP level was the primary direct-map estimate. Because these are different estimators, they were labeled explicitly and were not treated as identical measurement pipelines.

For each Monte Carlo realization, quantitative endpoints included level-wise \TtwoStar{} RMSE and mean absolute error, mean direct-map bias, median within-CEP voxel standard deviation, and the fraction of valid direct-map CEP voxels. Echo-decay fidelity was assessed using relative echo gain: $S_{\mathrm{recovered}}(\mathrm{TE})/S_{\mathrm{truth}}(\mathrm{TE})$, normalized by the corresponding first-echo gain. The RMS deviation of this normalized gain from 1 quantified echo-shape distortion. Mean image normalized RMSE across echoes was a secondary endpoint because image error does not directly measure quantitative relaxometry fidelity.

\subsection{Frozen Monte Carlo Validation and Statistical Analysis}
Twenty independent noise realizations were generated for each acceleration factor, yielding 100 final validation reconstructions per method. The representative $\Racc=8$ realization used for imaging figures was selected after all runs as the case with direct-map RMSE closest to the $\Racc=8$ median, rather than by visual appearance.

Results are reported as means with 95\% confidence intervals obtained from 2000 deterministic non-parametric bootstrap resamples of the 20 realization-level values. At $\Racc=8$, direct-model agreement was further evaluated over 100 level-by-realization observations (five CEP levels $\times$ 20 realizations) using ordinary least-squares slope/intercept, Lin's concordance correlation coefficient (CCC), mean bias, and Bland--Altman 95\% limits of agreement. No null-hypothesis significance testing was used for method ranking; interpretation emphasized effect sizes, confidence intervals, and estimator-specific failure modes.

\begin{table}[htbp]
\centering
\caption{Frozen synthetic acquisition and reconstruction design.}
\small
\begin{tabular}{@{}p{0.31\linewidth}p{0.62\linewidth}@{}}
\toprule
Component & Final frozen setting \\
\midrule
Matrix / geometry & $96 \times 96$ sagittal stylized lumbar phantom; five CEP levels \\
CEP ground truth & L1--L2 to L5--S1: 24, 22, 20, 18, 16 ms \\
UTE timing & TR 32 ms; flip angle $19^{\circ}$; TE 0.244, 5.2, 10.2, 15.2, 20.2, 25.2 ms \\
Receive array & 8 synthetic coils \\
Radial sampling & $\Racc=1/2/4/6/8$: 302/151/76/50/38 spokes \\
Final Monte Carlo & 20 independent noise realizations per $\Racc$; 100 total \\
CG-SENSE & Unregularized multi-coil least-squares reconstruction \\
Joint-FISTA & Shared multi-echo Haar-wavelet prior; frozen threshold $5 \times 10^{-5}$ \\
Direct model & Joint $\rho$/\RtwoStar{} nonlinear inversion; 75 iterations \\
Final regularization & $\rho$ wavelet threshold $=0$; \RtwoStar{} edge-aware $\beta=0$ \\
\bottomrule
\end{tabular}
\end{table}

\section{Results}
\subsection{Direct \TtwoStar{} Accuracy Degraded Progressively with Acceleration}
All 100 frozen Monte Carlo realizations completed successfully. Direct model-map level-wise \TtwoStar{} RMSE increased monotonically with radial undersampling: 0.172 ms (95\% CI, 0.153--0.191) at $\Racc=1$; 0.284 ms (0.260--0.306) at $\Racc=2$; 0.800 ms (0.766--0.836) at $\Racc=4$; 1.472 ms (1.386--1.557) at $\Racc=6$; and 2.256 ms (2.182--2.331) at $\Racc=8$. Mean direct-model bias was $-0.120$, $-0.247$, $-0.728$, $-1.395$, and $-2.136$ ms, respectively, indicating increasing negative bias with acceleration.

Spatial parameter stability also deteriorated with undersampling. Median within-CEP voxel standard deviation increased from 2.175 ms at $\Racc=1$ to 3.108 ms at $\Racc=2$, 5.802 ms at $\Racc=4$, 7.402 ms at $\Racc=6$, and 7.884 ms at $\Racc=8$. The direct map nevertheless returned valid values for all synthetic CEP voxels across the final validation.

\subsection{Image Fidelity and Quantitative \TtwoStar{} Fidelity Ranked Methods Differently}
At $\Racc=4$, joint-FISTA produced the lowest mean image NRMSE (0.132; 95\% CI, 0.132--0.132), compared with 0.203 for model-predicted echoes and 0.255 for CG-SENSE. However, the direct model-map level-wise \TtwoStar{} RMSE was 0.800 ms (0.766--0.836), whereas CG-SENSE and joint-FISTA ROI-fit RMSEs were 1.448 ms (1.382--1.510) and 1.288 ms (1.253--1.325), respectively. These comparisons involve different estimators but demonstrate that the method producing the lowest image error did not necessarily produce the lowest quantitative \TtwoStar{} error.

The divergence became more pronounced at severe undersampling. At $\Racc=8$, joint-FISTA again had the lowest image NRMSE (0.162), followed by the direct model predicted echoes (0.262) and CG-SENSE (0.349). In contrast, CG-SENSE ROI fitting had the lowest regional \TtwoStar{} RMSE (1.501 ms; 95\% CI, 1.404--1.608), followed by the direct model map (2.256 ms; 2.182--2.331), model-predicted echo ROI fitting (3.654 ms; 3.591--3.716), and joint-FISTA ROI fitting (3.842 ms; 3.820--3.866). Thus, severe undersampling reversed the ranking suggested by image NRMSE.

\subsection{Echo-Decay Distortion and High-Acceleration Failure Mode}
Echo-shape error increased with acceleration for all methods but differed substantially by reconstruction strategy. At $\Racc=8$, the RMS deviation of normalized relative echo gain from unity was 0.0666 for CG-SENSE, 0.1601 for the direct model, and 0.1766 for joint-FISTA. The combination of low image NRMSE and larger echo-shape distortion in joint-FISTA illustrates how image-domain regularization can alter the relative TE-dependent signal required for relaxometry.

\begin{table}[htbp]
\centering
\caption{Final Monte Carlo quantitative performance across acceleration. Direct model-map and ROI-fit values are different estimators and should not be interpreted as identical measurement pipelines.}
\scriptsize
\setlength{\tabcolsep}{3pt}
\resizebox{\textwidth}{!}{%
\begin{tabular}{@{}ccccc@{}}
\toprule
$\Racc$ & Spokes & Direct map RMSE, ms (95\% CI) & CG ROI-fit RMSE, ms (95\% CI) & Joint-FISTA ROI-fit RMSE, ms (95\% CI) \\
\midrule
1 & 302 & 0.172 (0.153--0.191) & 0.178 (0.151--0.204) & 0.632 (0.606--0.657) \\
2 & 151 & 0.284 (0.260--0.306) & 0.422 (0.395--0.449) & 0.612 (0.584--0.638) \\
4 & 76  & 0.800 (0.766--0.836) & 1.448 (1.382--1.510) & 1.288 (1.253--1.325) \\
6 & 50  & 1.472 (1.386--1.557) & 1.735 (1.687--1.784) & 2.787 (2.755--2.817) \\
8 & 38  & 2.256 (2.182--2.331) & 1.501 (1.404--1.608) & 3.842 (3.820--3.866) \\
\bottomrule
\end{tabular}%
}

\vspace{0.5em}
\begin{tabular}{@{}cc@{}}
\toprule
$\Racc$ & Direct bias / voxel SD, ms \\
\midrule
1 & $-0.120$ / 2.175 \\
2 & $-0.247$ / 3.108 \\
4 & $-0.728$ / 5.802 \\
6 & $-1.395$ / 7.402 \\
8 & $-2.136$ / 7.884 \\
\bottomrule
\end{tabular}
\end{table}

\begin{table}[htbp]
\centering
\caption{Image fidelity and echo-shape fidelity at moderate and severe acceleration.}
\scriptsize
\resizebox{\textwidth}{!}{%
\begin{tabular}{@{}ccccccc@{}}
\toprule
$\Racc$ & CG image NRMSE & Joint-FISTA image NRMSE & Model image NRMSE & CG echo error & Joint-FISTA echo error & Model echo error \\
\midrule
4 & 0.2548 & 0.1321 & 0.2027 & 0.0518 & 0.0628 & 0.0720 \\
8 & 0.3491 & 0.1616 & 0.2617 & 0.0666 & 0.1766 & 0.1601 \\
\bottomrule
\end{tabular}%
}
\end{table}

Across the 100 level-by-realization direct-model observations at $\Racc=8$, agreement with ground truth was moderate (CCC, 0.723). The regression slope was 0.836 with an intercept of 1.137 ms. Mean bias was $-2.136$ ms, and the Bland--Altman 95\% limits were $-3.605$ to $-0.668$ ms. Because both limits were below zero, the high-acceleration error represented systematic underestimation rather than random scatter centered on the ground truth. Level-wise direct-model bias at $\Racc=8$ ranged from $-3.25$ ms at L1--L2 to $-1.51$ ms at L5--S1, with the largest within-CEP voxel dispersion in the upper levels.

\section{Discussion}
This frozen synthetic Monte Carlo study produced three main findings. First, direct model-based CEP \TtwoStar{} mapping was quantitatively accurate at low and moderate radial acceleration, with a level-wise RMSE below 1 ms through $\Racc=4$. Second, reconstruction methods were ranked differently by image NRMSE and by the final \TtwoStar{} biomarker. Joint-FISTA consistently produced the lowest image NRMSE but developed larger quantitative error and echo-shape distortion at higher acceleration. Third, severe $\Racc=8$ undersampling exceeded the stable operating range of the current direct model, producing systematic \TtwoStar{} underestimation, reduced concordance, and large within-CEP voxel dispersion.

The divergence between image fidelity and biomarker fidelity is the central methodological result. Sparse image-domain regularization is designed to suppress aliasing and noise in reconstructed images. Relaxometry, however, depends on ratios between signals acquired at different echo times. If regularization shrinks or redistributes signal differently across echoes, the resulting image series can appear cleaner while the fitted decay slope is biased. The $\Racc=8$ results demonstrate this explicitly: joint-FISTA had approximately half the image NRMSE of CG-SENSE yet more than twice the regional \TtwoStar{} RMSE. For quantitative MRI, reconstruction should therefore be validated against the final biological parameter rather than using image quality alone as a surrogate endpoint.

Direct model-based reconstruction was introduced to constrain the echo series to a physically consistent exponential decay. This approach improved level-mean \TtwoStar{} recovery at moderate undersampling and follows prior nonlinear inverse methods that estimate relaxation maps directly from undersampled data \cite{block2009,sumpf2011,sumpf2014}. Nevertheless, the simple signal model was not sufficient at $\Racc=8$. The direct map became noisy and systematically underestimated \TtwoStar{}. Several mechanisms may contribute. Strong undersampling leaves insufficient data to estimate a spatially varying \RtwoStar{} field robustly; averaging a heterogeneous reconstructed parameter field also yields a mixture of exponentials at the ROI level. In addition, the present forward model is intentionally idealized and does not include off-resonance, gradient delays, actual 3D Cones/DIR-UTE preparation, fat-water interactions, or scanner-specific trajectory errors.

The behavior of CG-SENSE at $\Racc=8$ is also informative. Although CG-SENSE had the worst image NRMSE of the three iterative methods, its ROI-fit \TtwoStar{} RMSE and echo-shape fidelity were better than those of joint-FISTA and the direct model. In this synthetic experiment, spatial artifacts can partially average within the known CEP ROI while preserving relative decay across TE. This does not imply that CG-SENSE is preferable for clinical CEP imaging, where the true segmentation is unknown and streaking may compromise localization; rather, it emphasizes that regional parameter estimation and spatial image quality are distinct endpoints.

The results suggest $\Racc=4$ as a practical target for the next translational experiment. At this acceleration, the direct model-map RMSE was 0.800 ms with a narrow bootstrap confidence interval, while the experiment still showed substantial acceleration relative to the 302-spoke reference. We deliberately interpret $\Racc=6$ as a borderline stress test and $\Racc=8$ as a failure-mode condition rather than attempting to present all accelerations as successful. Such a bounded conclusion is more useful for sequence development because it defines where scanner validation should begin.

The lumbar CEP is particularly challenging because it is thin, orientation dependent, and adjacent to tissues with very different susceptibility and fat content. Fields et al. showed that CEP \TtwoStar{} at 3 T changes markedly with orientation relative to $B_0$ \cite{fields2015}. The present synthetic study held orientation fixed, meaning that the reported reconstruction errors should not be interpreted as total in vivo measurement error. Likewise, the predefined synthetic masks remove the segmentation uncertainty that would be present in human imaging. Prior UTE studies demonstrate that CEP morphology can be visualized and related to disc degeneration in cadaveric and clinical settings \cite{bae2013,finkenstaedt2023,ji2023,athertya2024}. Related quantitative UTE work in the musculoskeletal system has likewise emphasized noise robustness and the need for staged validation before clinical translation \cite{sedaghat2026}. Robust CEP mapping will require explicit control of orientation, partial volume, and sequence-specific contrast.

This study has limitations. The anatomy, tissue parameters, receive coils, and radial operator were synthetic. The tissue values were literature-anchored but were not patient-specific, and the level-wise \TtwoStar{} range was used as a controlled test pattern rather than as a diagnostic label. The encoding operator was matched between simulation and reconstruction, which is more favorable than real scanner conditions. We did not model motion, $B_0/B_1$ inhomogeneity, gradient delay, trajectory error, fat-water effects, partial volume, true 3D Cones sampling, or dual-inversion preparation. Noise was not calibrated to clinical repeatability. Finally, comparisons between the direct map and ROI-fit baselines involve different estimators and therefore should be interpreted as endpoint behavior, not as a definitive head-to-head clinical performance ranking.

The next phase should replace the matched synthetic operator with measured Siemens 3-T non-Cartesian trajectories and raw-data scaling, followed by coil-sensitivity estimation, gradient-delay and off-resonance correction, and physical phantom validation. A small repeatability experiment on a physical short-\TtwoStar{} phantom would substantially strengthen the quantitative claims before human testing. Once scanner fidelity is established, the same frozen endpoints---\TtwoStar{} bias, precision, echo-shape preservation, and acceleration dependence---can be applied to healthy-volunteer and patient data.

\section{Conclusions}
In a fully synthetic frozen Monte Carlo study of accelerated radial UTE MRI, image reconstruction quality did not reliably predict quantitative lumbar CEP \TtwoStar{} fidelity. Direct model-based reconstruction maintained sub-millisecond level-wise \TtwoStar{} RMSE through $\Racc=4$ but developed increasing negative bias and spatial instability at higher acceleration. Joint-FISTA provided superior image NRMSE yet greater \TtwoStar{} distortion at severe undersampling. These findings support endpoint-aware validation of accelerated quantitative MRI and motivate scanner/phantom translation beginning near $\Racc=4$ rather than further synthetic algorithm tuning.

\section*{Ethics Statement}
This work is a fully synthetic computational study. No human participants, animals, physical specimens, or identifiable clinical data were used. Institutional review board approval and informed consent were therefore not applicable.

\section*{Data and Code Availability}
The current manuscript draft describes a synthetic computational framework. Final repository, code-availability, and data-sharing statements should be completed by the author group before submission.

\clearpage
\begin{figure}[p]
\centering
\includegraphics[width=\textwidth]{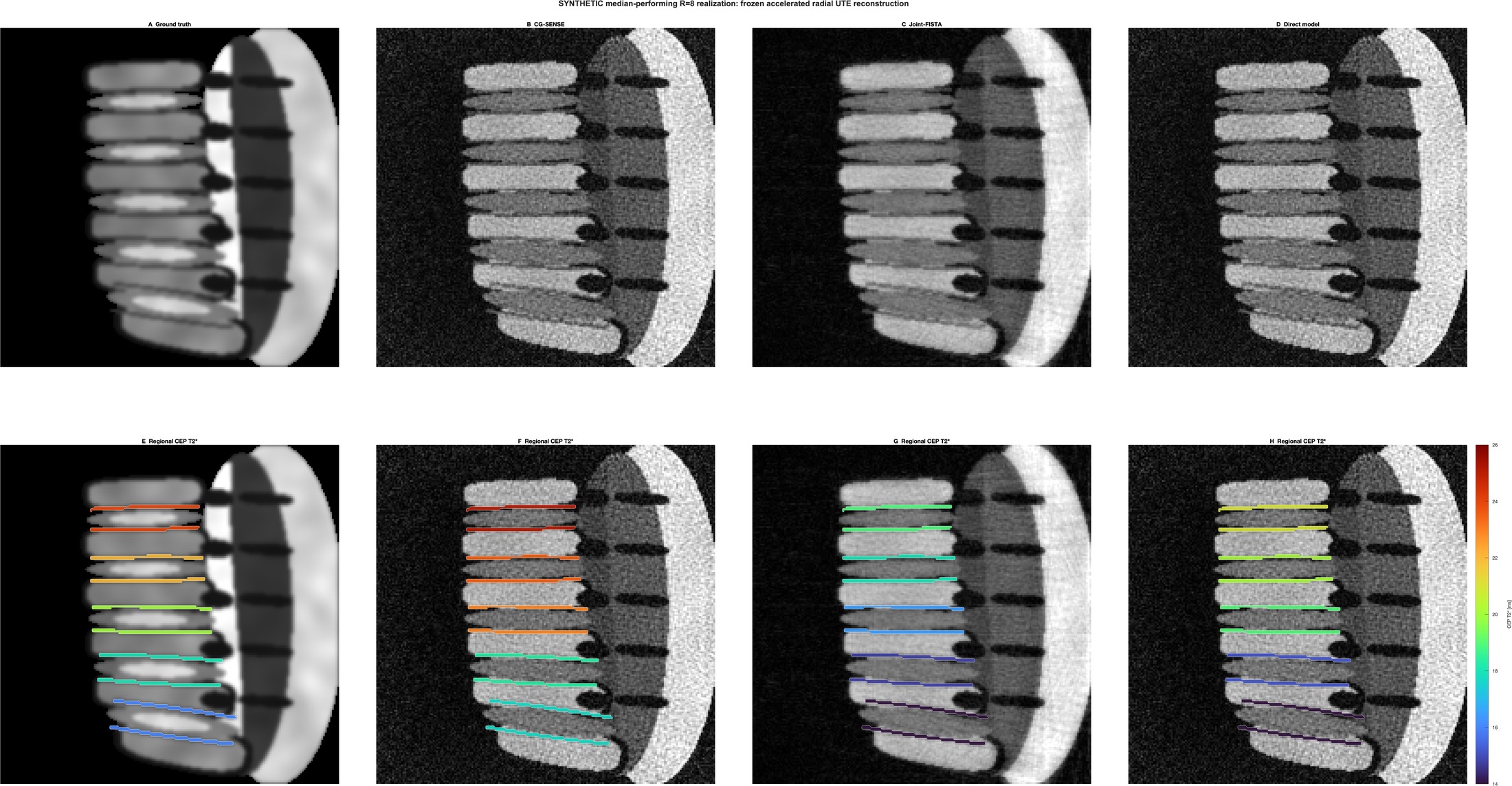}
\caption{Representative median-performing $\Racc=8$ realization. Top row: synthetic reference, CG-SENSE, joint-FISTA, and direct model-predicted first-echo magnitude images. Bottom row: corresponding regional CEP \TtwoStar{} visualization. The $\Racc=8$ case was selected by median direct-model RMSE, not visual appearance. The panel is a stress-test/failure-mode example and should not be interpreted as the preferred operating point.}
\end{figure}
\clearpage

\begin{figure}[p]
\centering
\includegraphics[width=\textwidth]{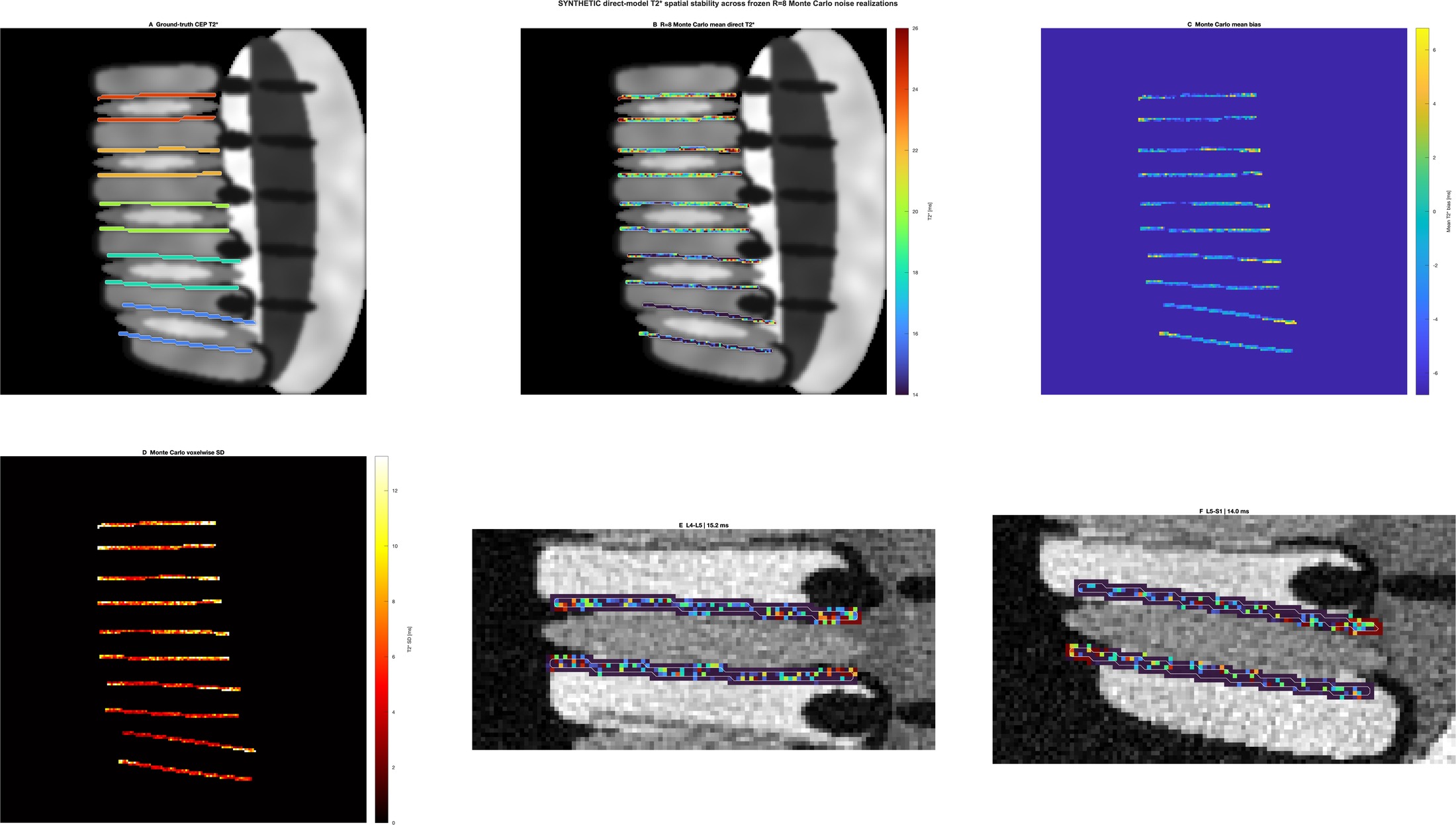}
\caption{Spatial stability of the direct model at $\Racc=8$ across 20 frozen noise realizations. Ground-truth CEP \TtwoStar{}, Monte Carlo mean direct map, mean bias, and voxelwise standard deviation illustrate systematic underestimation and high spatial variance under severe radial undersampling; lower-lumbar crops show representative regional maps.}
\end{figure}
\clearpage

\begin{figure}[p]
\centering
\includegraphics[width=\textwidth]{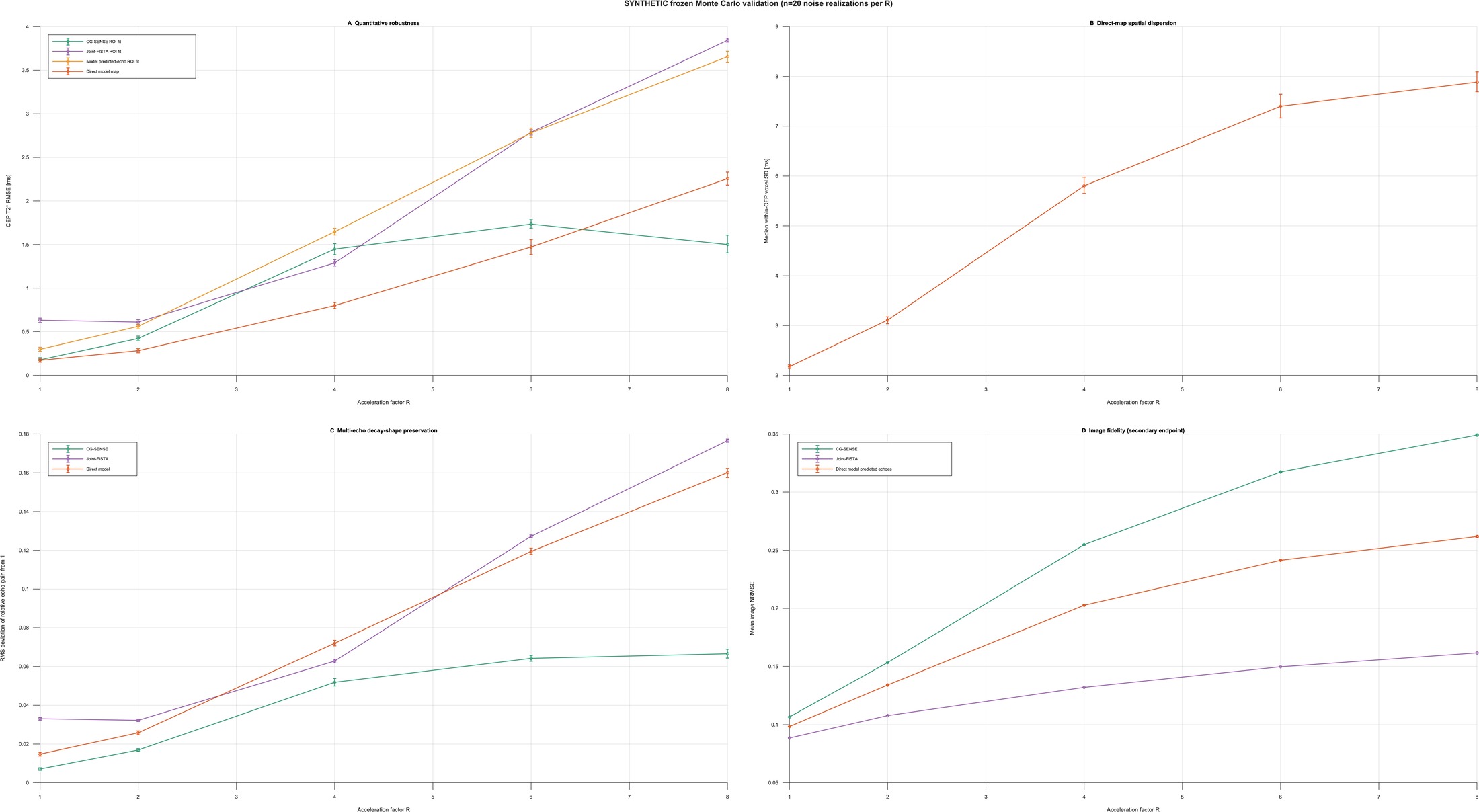}
\caption{Frozen Monte Carlo performance across $\Racc=1,2,4,6,$ and 8 (20 noise realizations per acceleration). Quantitative \TtwoStar{} error, direct-map dispersion, echo-shape error, and image NRMSE are shown with 95\% bootstrap confidence intervals. The key result is the divergence between image fidelity and quantitative biomarker fidelity as acceleration increases.}
\end{figure}
\clearpage

\begin{figure}[p]
\centering
\includegraphics[width=\textwidth]{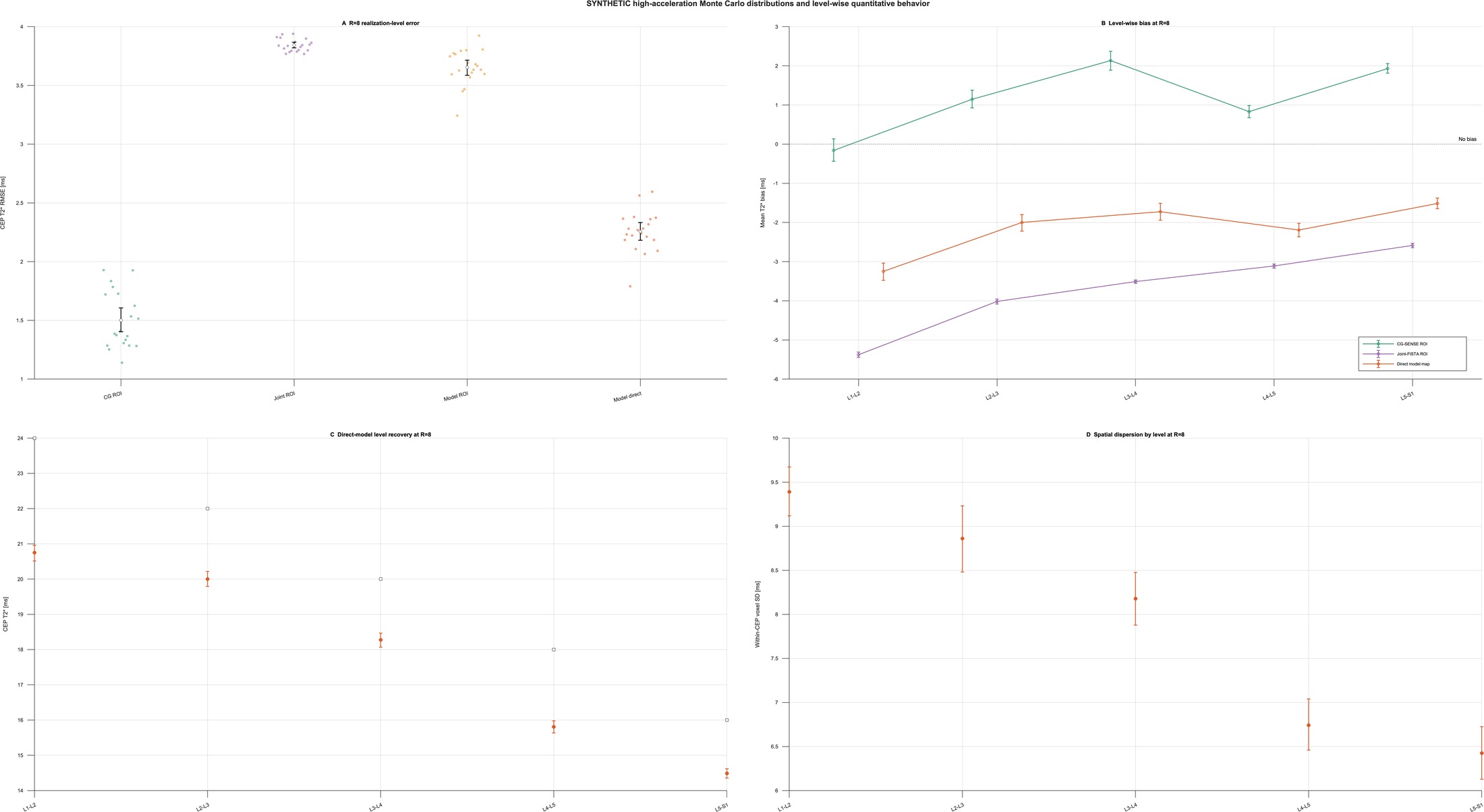}
\caption{High-acceleration $\Racc=8$ distributions and level-wise behavior. Realization-level \TtwoStar{} errors, level-wise biases, direct-model \TtwoStar{} recovery, and within-CEP dispersion demonstrate that the $\Racc=8$ failure mode is systematic and spatially heterogeneous rather than a single unfavorable realization.}
\end{figure}
\clearpage

\begin{figure}[p]
\centering
\includegraphics[width=\textwidth]{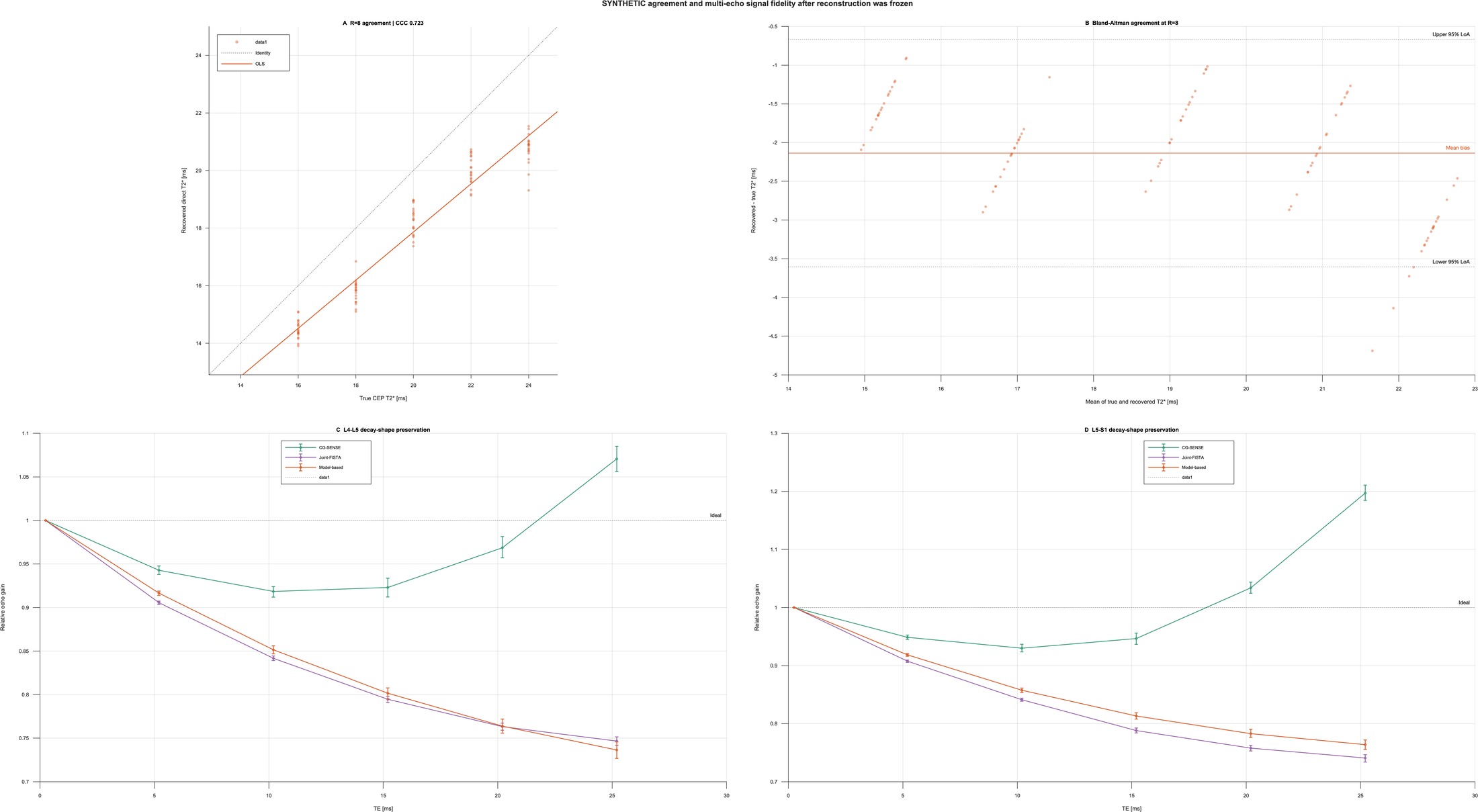}
\caption{$\Racc=8$ agreement and echo-decay fidelity. Direct-model estimates are compared with ground truth using identity-line regression and Bland--Altman analysis; multi-echo relative-gain curves demonstrate reconstruction-dependent distortion of the TE decay. At $\Racc=8$, the direct model had CCC 0.723 and mean bias $-2.136$ ms, with 95\% limits of agreement from $-3.605$ to $-0.668$ ms.}
\end{figure}

\end{document}